# New insights on the Dynamic Cellular Metabolism

Ildefonso M. De la Fuente [1,2]

1 Department of Cell Biology and Immunology, Institute of Parasitology and Biomedicine "López-Neyra", CSIC, Granada, Spain.
2 Department of Mathematics, University of the Basque Country, UPV/EHU, Leioa. Spain.

**Corresponding Author:**
Ildefonso Mtz. de la Fuente Mtz.
Instituto de Parasitología y Biomedicina "López-Neyra", CSIC,
Parque Tecnológico de Ciencias de la Salud
Avda. del Conocimiento s/n.
18016 Granada. ESPAÑA.
 E-Mail: mtpmadei@ehu.es
Tel.: +34- 958-18-16-21; Fax: +34- 958-18-16-32.

## Abstract

A large number of studies have shown the existence of metabolic covalent modifications in different molecular structures, able to store biochemical information that is not encoded by the DNA. Some of these covalent mark patterns can be transmitted across generations (epigenetic changes). Recently, the emergence of Hopfield-like attractor dynamics has been observed in the self-organized enzymatic networks, which have the capacity to store functional catalytic patterns that can be correctly recovered by the specific input stimuli. The Hopfield-like metabolic dynamics are stable and can be maintained as a long-term biochemical memory. In addition, specific molecular information can be transferred from the functional dynamics of the metabolic networks to the enzymatic activity involved in the covalent post-translational modulation so that determined functional memory can be embedded in multiple stable molecular marks. Both the metabolic dynamics governed by Hopfield-type attractors (functional processes) and the enzymatic covalent modifications of determined molecules (structural dynamic processes) seem to represent the two stages of the dynamical memory of cellular metabolism (metabolic memory). Epigenetic processes appear to be the structural manifestation of this cellular metabolic memory. Here, a new framework for molecular information storage in the cell is presented, which is characterized by two functionally and molecularly interrelated systems: a dynamic, flexible and adaptive system (metabolic memory) and an essentially conservative system (genetic memory). The molecular information of both systems seems to coordinate the physiological development of the whole cell.

# 1. Enzymatic self-organizations

## 1.1 Multienzymatic complexes

Living cells are essentially highly evolved dynamic metabolic reactors, in which the more complex known molecules are synthesized and destroyed by means of interrelated enzymatic processes, densely integrated in modular networks that shape one of the most complex dynamic systems in nature [De la Fuente, 2014].

From a functional point of view, the enzymes are the main molecules of this amazing biochemical reactor. They are responsible for almost all the biomolecular transformations, which, when globally considered, are called cellular metabolism. Most enzymes are proteins, but a few RNA molecules called ribozymes, also manifest catalytic activity.

How the enzymes are organized under the complex conditions prevailing inside the cell is a crucial issue for the understanding of the fundamental functional architecture of cellular life. For many years, it has been assumed that enzymes work in an isolated way. However, intensive studies of protein-protein interactions have shown that the internal cellular medium is an assembly of supra-molecular protein complexes [Gavin and Superti-Furga, 2003; Segura et al., 2010; Völkel et al., 2010; Ramisetty and Washburn, 2011], where homologous or heterologous protein-protein interactions are the rule [Pang et al., 2008]. Concretely, analyses of the proteome of *Saccharomyces cerevisiae* have shown that at least 83% of proteins form complexes comprised by two to eighty-three proteins [Gavin et al., 2002]. This kind of associative organization occurs in all sorts of cells, both eukaryotes and prokaryotes [Bobik, 2006; Ho et al., 2002; Ito et al., 2001; Sutter et al., 2008; Uetz et al., 2000; Yeates et al., 2008].

Multienzymatic associations forming complexes may allow for substrate channelling, i.e. direct transfer of intermediate metabolites from the active site of one enzyme to the next without prior diffusion into the bulk medium. Metabolic channelling decreases the transit time of reaction substrates, thus increasing the speed and efficiency of catalysis by preventing delays due to diffusion of reaction intermediates and their eventual loss [Clegg and Jackson, 1990; Ovadi and Srere, 2000; Jorgensen et al., 2005; Jovanović et al., 2007]. Substrate channelling may also occur within channels or on the electrostatic surface of enzymes belonging to complexes [Milani et al., 2003; Ishikawa et al., 2004; Beg et al., 2007].

Additionally, reversible interactions between multienzyme complexes and structural proteins or membranes occur frequently in eukaryotic cells leading to the emergence of metabolic microcompartments [Verkman, 2002; Lunn, 2007; Monge et al., 2008; Monge et al., 2009; Saks et al., 2009]. For instance, the nuclear processes such as DNA replication, DNA repair and transcription of RNAs are organized in dynamic microenvironments that are functionally connected to the ubiquitin–proteasome system [Zhao et al., 2009].

Microcompartmentation also happen in prokaryotes but in this case they consist of protein "shells" composed of thousands of protein subunits [Yeates et al., 2007; Fan et al., 2010]. The dynamics of molecular processes that intervene in microcompartmentation and its maintenance remain unclear.

The concept of multienzymatic associations was first conceived in 1970 by A. M. Kuzin [Kuzin, 1970] and adopted in 1972 by P. A. Srere [Srere, 1972; Srere, 1985] who suggested the use of the term metabolon to refer to an organized multienzyme complex belonging to a metabolic sequence.

A wide variety of studies point to the importance of multienzymatic associations in almost all processes occurring in cells [Veliky et al., 2007; Pflieger et al., 2011] e.g., the purinosome [Zhao et al., 2013], the aminoacyl-tRNA synthetase complex [Bhaskaran and Perona, 2011], the urea cycle [Cheung et al., 1989], the ATP synthasome [Clémençon, 2012], respiratory chain supercomplex [Schägger, 2002; Stuart, 2008; Bultema et al., 2009], the Benson-Calvin cycle [Suss et al., 1993; Graciet et al., 2004], the replisome [Murthy and Pasupathy, 1994], the amino acid catabolism complex [Islam et al., 2007], the cellulosome [Bayer et al., 2004], the sporopollenin biosynthesis complex [Lallemand et al., 2013], the spliceosome [Will and Lührmann, 2011], the 21 S complex of enzymes for DNA synthesis [Li et al., 1993], the TRAMP complex [Jia et al., 2012], the glycogen biosynthesis complex [Wilson et al., 2010], the degradosome [Carpousis, 2002], the vault ribonucleoprotein complex [Zheng et al., 2005], the proteasome [Murata et al., 2009], the fatty acid oxidation complex [Binstock and Schulz, 1981; Parker and Engel, 2000], the ribosome [Lin et al., 2013], the glycolytic enzymes associate [Waingeh et al., 2006; Graham et al., 2007; Forlemu et al., 2011], the protein kinase complexes [Rohila et al., 2009], the photosystem I [Fromme and Mathis, 2004], the Krebs cycle [Barnes and Weitzman, 1986; Beeckmans et al., 1990; Mitchell, 1996], the COP9 signalosome complex [Schwechheimer, 2004], the ski complex [Halbach et al., 2013], the exosome [Makino et al., 2013], the dystrophin-associated protein complex [Ehmsen et al., 2002], the arginine biosynthesis complex [Abadjieva et al., 2001], and the mitotic and cytokinetic protein complexes [D'Avino et al., 2009].

Recently, analysis of conserved sequences in biochemical reactions has shown that the cellular metabolism contains basic functional units which tend to correspond to traditional metabolic pathways. These functional multi-enzymatic units seem to be catalytic building blocks of the cellular metabolism [Kanehisa, 2013; Muto et al., 2013].

Another kind of functional catalytic association is the interactome which refers to protein–protein interaction networks (unlike metabolic pathways, that represent a sequence of molecular interactions leading to a final product) [Van Leene et al., 2007; Bonetta, 2010; Gautam et al., 2012; Sharma et al., 2013].

**1. 2 Temporal self-organization of multienzymatic processes**

The enzymatic organization at the molecular level presents another relevant characteristic: the emergence of dissipative catalytic structures.

Experimental observations have shown that the enzymes shape functional catalytic associations in which molecular rhythms and quasi-steady states may spontaneously emerge far from thermodynamic equilibrium. In fact, under cellular conditions, the time evolution of practically all metabolite concentrations is subjected to marked variations, presenting transitions between quasi-steady states and oscillatory behaviors [De la Fuente, 2014]. In the quasi-stationary states the concentration of any given metabolite drifts over time in a non-oscillatory way. Not long ago, the use of nanobiosensors has contributed with real-time measurements of determinate intracellular metabolites in intact cells, which show complex oscillations and quasi-steady states, but importantly, these patterns are never constant [Ozalp et al., 2010].

Metabolic rhythms constitute a genuine manifestation of the dissipative self-organization in the enzymatic activities of cell. In general terms, self-organization can be defined as the spontaneous emergence of macroscopic non-equilibrium dynamic structures, as a result of collective behavior of elements, interacting nonlinearly with each other, to generate a system that increases its structural and functional complexity driven by energy dissipation [Halley and Winkler, 2008; Misteli, 2009; De la Fuente, 2014].

An important ingredient of metabolic self-organization is given by nonlinear interaction between metabolites and enzymes, involving autocatalysis, feed-back processes, and allosteric regulation among others.

It is well established that non-equilibrium states can be a source of order in the sense that the irreversible processes may lead to a new type of dynamic state in which the system becomes ordered in space and time. In fact, the non-linearity associated to the irreversible enzymatic reactions seems to be an essential mechanism which may allow the dissipative metabolic organization far from thermodynamic equilibrium [Goldbeter, 2002; Goldbeter, 2007].

Self-organization is based on the concept of dissipative structures, and its theoretical roots can be traced back to the Nobel Prize in Chemistry Ilya Prigogine [Nicolis and Prigogine, 1977].

There exist different kinds of metabolic dissipative structures in cellular conditions, mainly oscillations in the metabolite concentrations, molecular circadian rhythms and spatial biochemical waves.

During the last four decades, extensive studies of biochemical dynamics both in prokaryotic and eukaryotic cells have revealed that oscillations (temporal self-organizations) exist in most of the fundamental metabolic processes. For instance, specific molecular oscillations were reported to occur in: free fatty acids [Getty-Kaushik et al., 2005], NAD(P)H concentration [Rosenspire et al., 2001], biosynthesis of phospholipids [Marquez et al., 2004], cyclic AMP concentration [Holz et al., 2008], ATP [Ainscow et al., 2002] and other adenine nucleotide levels [Zhaojun et al., 2004], intracellular glutathione concentration [Lloyd and Murray, 2005], actin polymerization [Rengan and Omann, 1999], ERK/MAPK metabolism [Shankaran et al., 2009], mRNA levels [Zhaojun et al., 2004], intracellular free amino acid pools [Hans et al., 2003], cytokinins [Hartig and Beck, 2005], cyclins [Hungerbuehler et al., 2007], transcription of cyclins [Shaul et al., 1996], gene expression [Chabot et al., 2007; Tian et al., 2005; Tonozuka et al., 2001; Klevecz et al., 2004], microtubule polymerization [Lange et al., 1988], membrane receptor activities [Placantonakis and Welsh, 2001], membrane potential [De Forest and Wheeler, 1999], intracellular pH [Sánchez-Armáss et al., 2006], respiratory metabolism [Lloyd et al., 2002], glycolysis [Dano et al., 1999], intracellular calcium concentration [Ishii et al., 2006], metabolism of carbohydrates [Jules et al., 2005], beta-oxidation of fatty acids [Getty et al., 2000], metabolism of mRNA [Klevecz and Murray, 2001], tRNA [Brodsky et al.,1992], proteolysis [Kindzelskii et al., 1998], urea cycle [Fuentes et al., 1994], Krebs cycle [Wittmann et al., 2005], mitochondrial metabolic processes [Aon et al., 2008], nuclear translocation of the transcription factor [Garmendia-Torres et al., 2007], amino acid transports [Barril and Potter, 1968], peroxidase-oxidase reactions [Møller et al., 1998], protein kinase activities [Chiam and Rajagopal, 2007] and photosynthetic reactions [Smrcinová et al., 1998].

Experimental observations performed in *S. cerevisae* during continuous culture have shown that most of transcriptome and the metabolome exhibit oscillatory dynamics [Klevecz et al., 2004; Murray et al., 2007]. In some studies, it has been observed that at least 60% of all gene expression exhibit oscillations with an approximate period of 300 min [Tu et al., 2005]. Moreover, other studies have shown that the entire transcriptome exhibits low-amplitude oscillatory behavior [Lloyd and Murray, 2006] and this phenomenon has been described as a genome-wide oscillation [Klevecz et al., 2004, Lloyd and Murray, 2005; Oliva et al., 2005; Tu et al., 2005; Lloyd and Murray; 2006, Murray et al., 2007].

Evidence that cells exhibit multi-oscillatory metabolic processes with fractal properties has also been reported. This dynamic behavior appears to be consistent with scale-free dynamics spanning a wide range of frequencies of at least three orders of magnitude [Aon et al., 2008]. The temporal organization of the metabolic processes in terms of rhythmic phenomena spans periods ranging from milliseconds [Aon et al., 2006], to seconds [Roussel et al., 2006], minutes [Berridge and Galione, 1988; Chance et al., 1973] and hours [Brodsky, 2006].

Oscillatory behavior from simple patterns to complex temporal structures [MacDonald et al., 2003], including bursting oscillations, with one large spike and series of secondary oscillations [Dekhuijzen and Bagust, 1996], have often been observed.

Multiple autonomous oscillatory behaviors emerge simultaneously in the cell, so that the whole metabolism resembles a complex multi-oscillator super-system [Lloyd and Murray, 2005; Lloyd and Murray, 2006; Murray et al., 2007; Aon et al., 2008; Sasidharan et al., 2012; Amariei et al., 2014; De la Fuente et al., 2014].

Numerous mathematical studies on temporal metabolic rhythms have contributed to a better understanding of the self-organized enzymatic functionality in cellular conditions [Boiteux et al., 1980; Tyson, 1991; De la Fuente et al., 1996; Heinrich and Schuster 1996; De la Fuente et al., 1999; Novak et al., 2007; Zhou et al., 2009; Goldbeter, 2013].

The dissipative self-organization at the enzymatic level presents other kinds of emergent dynamic structures as, for instance, circadian rhythms [Schibler and Sassone-Corsi, 2002; Cardone et al., 2005; De la Fuente, 2010] and spatial biochemical waves [Petty and Kindzelskii, 2001; Bernardinelli et al., 2004; Asano et al., 2008; De la Fuente, 2010].

Metabolic rhythms constitute one of the most genuine properties of multienzymatic dynamics. The conditions required for the emergence and sustainability of these rhythms, and how they are regulated, represent a biological problem of the highest significance. However, in spite of its physiological importance, many aspects of the cellular dissipative structures are still poorly understood and thus deserve further attention.

## 1. 3 Dissipative multienzymatic complexes: Metabolic Subsystems

Self-organized multienzymatic complexes can be viewed as dissipative structures in which molecular rhythms and functional integrative processes can emerge, increasing the efficiency and control of the biochemical reactions involved [De la Fuente, 2010; De la Fuente and Cortes, 2012]. This concept was conceived in 1999, being suggested the use of the term *Metabolic Subsystem* to refer to a dissipativelly structured enzymatic set in which molecular oscillations and steady state patterns may emerge spontaneously [De la Fuente et al., 1999].

Thermodynamically, metabolic subsystems represent advantageous biochemical organizations, forming unique, well-defined autonomous dynamic systems, in which the catalytic activity is autonomous with respect to the other enzymatic associations. Therefore, each self-organized enzymatic set shapes a catalytic entity as a whole and carry out its activities relatively independently of others [De la Fuente, 2010; De la Fuente, 2014].

The presence of some regulatory proteins (enzymes of both allosteric and covalent modulation) in each metabolic subsystem makes possible the sophisticated interconnections among them. As a result of the variable combinations in the different input conditions (substrate fluxes, allosteric signals, specific post-translational modulations, etc.) a rich variety of dynamic patterns emerge in each metabolic subsystem, which correspond to distinct catalytic activity regimes [De la Fuente et al., 2013; De la Fuente, 2014]. This huge variety of catalytic patterns is a sine qua non condition for the complex information properties (with capacity to store functional catalytic patterns) that can emerge in the cellular metabolic networks (see Section Three for more details).

Metabolic subsystems represent highly efficient nanomachines that can perform autonomous and discrete biological functions, allowing the structural and functional dynamic organization of the cell.

Associations between metabolic subsystems can form higher level complex molecular organizations, as for example it occurs with Intracellular Energetic Units (ICEU) [Saks et al., 2006] and synaptosomes [Monge et al., 2008].

In essence, metabolic subsystems are self-organized biochemical machines which seem to constitute a fundamental and elemental catalytic unit of the cellular biochemical reactor [De la Fuente, 2014].

## 2. Functional organization of enzymes in metabolic networks

**2.1 Metabolic functionality**

All physiological processes rely on metabolic functionality, which is a key element to understanding the dynamic principles of cellular life at all levels of its biochemical organization [Noble et al., 2014].

Enzymatic activity depends on different factors, mainly substrate concentration, temperature, pH, enzyme concentration and the presence of any inhibitors or activators.

When an enzymatic system operates sufficiently far from equilibrium, oscillatory catalytic patterns can emerge. Such dynamic behaviors find their roots in the non-linear regulatory processes, e.g., feedback loops, cooperativity and stoichiometric autocatalysis [Goldbeter, 2002].

In addition, due to the collective interactions under cellular conditions, substrate fluxes and regulatory signals shape complex topological structures in which the metabolite concentration is continuously changing over time because of the dynamics that emerge in the catalytic networks [De la Fuente, 2014]. These collective enzymatic connectivities make the rate at which enzymatic reactions proceed highly complex.

As a consequence, the functional architecture of the cell is the result of all these different factors, particularly of the complex dynamic interactions between enzymatic processes integrated in the biochemical networks.

In brief, metabolic functionality at a systemic level prominently depends on the collective architecture (structural and functional) of the catalytic reactions, which is defined by the two essential principles that link the different modes of enzymatic connectivity: the *metabolic segregation* and the *metabolic integration*.

**2. 2 Metabolic segregation**

The separation of metabolic tasks and catalytic roles demands that the enzymes with correlated biochemical activities are grouped together. As a consequence, the entire metabolism is functionally segregated at multiple levels of specialization (e.g., the signal transduction, the oxidative phosphorylation, the lysosomal metabolic activity, etc.), originating different kinds of specific modular structures.

A metabolic module can be considered as a discrete functional entity, which performs relatively autonomous processes with specific and coherent dynamic activities [Hartwell et al., 1999].

A wide variety of studies have shown that the cellular metabolic networks are organized in a specialized modular fashion, which constitutes a defining characteristic of the metabolic organization in the cell [Ravasz et al., 2002; Ravasz, 2009; Nurse, 2008; Hao et al., 2012; Kaltenbach and Stelling, 2012; Geryk and Slanina, 2013].

Modular segregations have been observed in all main physiological processes, e.g., the enzymatic organization of metabolic pathways [Muto et al., 2013], the chaperone activities [Korcsmáros et al., 2007], the chemotaxis [Postma et al., 2004; Shimizu et al., 2010], the apoptosis processes [Harrington et al., 2008], the cell cycle [Boruc et al., 2010; Hsu et al., 2011], the Golgi apparatus [Nakamura et al., 2012], and the kinetochore organization [Petrovic et al., 2014]. Specific modular metabolic networks also participate in the transcriptional system such as the miRNA regulatory networks [Gennarino et al., 2012], the transcription factors networks [Neph et al., 2012], the RNA polymerase complexes [Schubert, 2014], and the mRNA dynamics [Vlasova-St Louis and Bohjanen, 2011]. Metabolic subsystems, i.e., self-organized multienzymatic associations, can be considered as elemental catalytic modules of the cell [De la Fuente et al., 2008, De la Fuente, 2010; De la Fuente, 2014].

Metabolic networks may also contain canonical motifs, which represent recurrent and statistically significant sub-graphs or patterns [Milo et al., 2002; Kashtan and Alon, 2005; Alon, 2007; Masoudi-Nejad et al., 2012]. These sub-graphs that repeat themselves in a specific network are defined by a determinate pattern of interactions, which may reflect particular functional properties but, in contrast with the modules, motifs do not function in isolation [Lacroix et al., 2006].

The functional role played by the specialized metabolic networks has been defined largely by its structural connections. However, at least three main kinds of metabolic connectivity must be considered: the *structural connectivity*, which is formed by structural molecular links such as substrate fluxes and regulatory signals, the *functional connectivity*, which describes statistical dependencies (non-causal) between the activity patterns, and the *effective connectivity*, which rests on the causal effects between different biochemical activities [De la Fuente et al., 2011; De la Fuente et al., 2013].

A number of information-based dynamic tools for the functionality and correlations between biochemical processes have been proposed. However, functional correlations (*functional connectivity*) do not imply *effective connectivity* because most synchronization measures do not distinguish between causal and non-causal interactions.

In recent studies, Transfer Entropy (TE) has been proposed to be a rigorous, robust and self-consistent method for the causal quantification of the functional information flow between nonlinear processes (*effective connectivity*). In fact, TE allows to quantify the reduction in uncertainty that one variable has on its own future when adding another, allowing for a calculation of the functional influence in terms of effective connectivity between two variables [Schreiber, 2000].

The quantitative study of the enzymatic effective connectivity can be considered at different biochemical organization levels, e.g., in a biochemical pathway [De la Fuente and Cortes, 2012] and in a metabolic network [De la Fuente et al., 2011]. For example, in order to investigate the functional structure (both functional and effective connectivity) in a large number of self-organized enzymes, the effective connectivity has been analyzed by means of TE in dissipative metabolic networks (DMNs) under different external conditions [De la Fuente et al., 2011].

Essentially, a DMN is a set of interconnected nodes, each one correspond to a specific metabolic subsystem, which can be considered as elemental self-organized catalytic modules of the cell (see Section One for more details). The DMN´s nodes are interconnected through a complex topological structure (structural connectivity) of substrate fluxes and different regulatory mechanisms: activatory and inhibitory allosteric modulations, and post-translational modulations (PTM or covalent modulation in short). In agreement with experimental observations, the emergent output activity of each enzymatic subsystem in the DMN´s can be either oscillatory or steady state and comprises an infinite number of distinct activity regimes [De la Fuente et al., 2008; De la Fuente et al., 2010; De la Fuente and Cortes, 2013].

The TE analysis in the DMN showed that in addition to the network's topological structure, characterized by the specific location of enzymatic subsystems, molecular substrate fluxes and regulatory signals, there is another functional structure of biomolecular information flows which is modular, dynamical and able to modify the catalytic activities of all the enzymatic sets. Specifically, certain self-organized enzymatic sets are spontaneously clustered forming functional metabolic sub-networks. Some modules are preserved under different external conditions while others are preserved but with the inverted fluxes and yet other modules emerge only under specific external perturbations. These modules work efficiently: very concrete enzymatic subsystems are connected with others by means of effective information flows allowing high coordination and precise catalytic regulations. The dynamic changes between the functional modules correspond to metabolic switches, which allow for critical transitions in the enzymatic activities. The coordinated modules by effective connectivity and the functional switches seem to be important regulatory mechanisms that are able to modify the catalytic activity of all the enzymatic sets [De la Fuente et al., 2011].

**2.3 Metabolic integration**

In addition to the functional segregation, the cellular enzymatic processes exhibit another fundamental principle of metabolic connectivity: the functional integration.

Catalytic systems organized into distinct modular networks seem to be integrated, emerging systemic responses from the concerted actions of the specialized metabolic processes [Almaas et al., 2004; Almaas et al., 2005; Yoon et al., 2007; Buescher et al., 2012; San Roman et al., 2014].

Whilst functional segregation expresses the relative autonomy of specialized biochemical networks, metabolic integration is a complementary principle that refers to significant deviations from autonomy of the modular networks, so that the collective catalytic behavior generates coherent global patterns and molecular information that is simultaneously highly diversified and highly integrated [De la Fuente et al., 2011].

Different mechanisms are responsible for establishing functional metabolic integration. Studies with TE in DMNs have shown that the integration of specialized biochemical areas is mediated by effective connectivity processes which lead to the emergence of a systemic functional structure that encompasses the whole metabolic system. This global structure is characterized by a small set of different enzymatic processes always locked into active states (metabolic core) while the rest of the catalytic reactions present on-off dynamics [De la Fuente et al., 2011].

Already in 1999, in an exhaustive analysis with several millions of different DMNs, the emergence of this functional structure at global level was observed for the first time [De la Fuente et al., 1999]. Afterwards, different studies implementing flux balance analysis in experimental data added new evidence of this systemic metabolic structure in the cell by which a small set of enzymatic reactions belonging to different metabolic processes remain active under all investigated growth conditions (metabolic core), and the remaining enzymatic reactions stay only intermittently active. The emergence of the global catalytic structure was verified for *E. coli*, *H. pylori*, and *S. cerevisiae* [Almaas et al., 2004; Almaas et al., 2005]. In these studies, the reactions present in the metabolic core were also characterized, which seemed to be essential for biomass formation and for ensuring optimal metabolic performance. Likewise, the percentage of enzymatic reactions belonging to the core turned out to be small: 36.2% of all different reactions in *H. pylori*, 11.9% in *E. coli* and 2.8% in *S. cerevisiae*.

Similar results about the size of the metabolic cores have been obtained in extensive numerical studies with DMNs, in which the sizes ranged between 44.6% and 0.7% and their distribution is skewed toward low values since 71% of the studied networks show sizes that range between 7.6% and 0.7%, with a mean of 2% [De la Fuente et al., 2009].

In addition to the emergence of a global metabolic structure [De la Fuente et al., 2008], metabolic analysis implementing flux balance analysis and other studies on effective connectivity in DMNs have shown that two main dynamic mechanisms are involved in the systemic metabolic self-regulation: *flux plasticity* and *structural plasticity* [Almaas et al., 2004, Almaas et al., 2005; De la Fuente et al., 2011]. Flux plasticity represents changes in the intensity of the substrate fluxes. Structural plasticity results in persistent changes in the dynamic structure conformed by the catalytic processes that present *on-off* and *on* changing states.

Another mechanism involved in the functional metabolic integration is the energy.

Living cells require a permanent generation of energy flow to keep the functionality of their complex metabolic structures, which integrate a large ensemble of modular networks. Practically all bioenergetic processes are coupled with each other via adenosine nucleotides, which are consumed or regenerated by different enzymatic reactions. Moreover, the adenosine nucleotides also act as an allosteric control of numerous regulatory enzymes allowing changes in ATP, ADP and AMP levels to regulate the functional activity of those metabolic subsystems that exhibit this allosteric mechanism.

At a systemic level, the temporal dynamics for the concentrations of adenosine nucleotides seem to be determined by the Adenylate Energy System which represents the synthesis of ATP coupled to energy-consumption processes through a complex network of enzymatic reactions which interconvert ATP, ADP and AMP [De la Fuente et al., 2014].

Experimental studies have shown that the adenosine nucleotide dynamics present complex transitions across time, e.g., the use of nanobiosensors has shown real-time-resolved measurements of intracellular ATP in intact cells. In these experimental observations, the ATP concentration showed a rhythmic behavior and complex quasi steady-state dynamics with large variations over time but, importantly, the ATP concentration is never constant [Ozalp et al., 2010; Ytting et al., 2012].

The concentration levels of ATP, ADP and AMP roughly reflect the energetic status of the cell, and a determined systemic ratio between them was proposed by Atkinson as the adenylate energy charge (AEC) [Atkinson and Walton, 1967] which is a scalar index ranging between 0 and 1. When all adenine nucleotide pool is in form of AMP, the energy charge is zero, and the system is completely discharged (zero concentrations of ATP and ADP). With only ADP, the energy charge is 0.5. If all adenine nucleotide pool is in form of ATP, the AEC is 1.

Under growth conditions, the organisms seem to maintain their AEC within narrow physiological values (AEC values between 0.7 and 0.95), despite of extremely large fluctuations in the adenine nucleotide concentrations. This systemic physiological phenomenon, constitute a key property of the functional integration of metabolic processes in the cell [De la Fuente et al., 2014].

The main biological significance of the maintained AEC level appears to be the necessity of having the adenylate pool highly phosphorylated keeping the rate of adenylate energy production similar to the rate of adenylate energy expenditure [De la Fuente et al., 2014].

Intensive experimental studies have shown that the physiological AEC ratio is practically preserved in all unicellular organisms, both eukaryotes and prokaryotes. Besides, the ratio is also strictly fulfilled during all the metabolic transformations that occur during the cell cycle, which seems to be a manifestation of the integrative processes of the systemic metabolism [De la Fuente et al., 2014].

Functional specialization and integration are not exclusive; they are two dynamic complementary principles, and the interplay between them seems to be an essential element for the functional collective architecture of the cell systemic metabolism.

# 3. Metabolic networks and information properties

## 3.1 Experimental findings of biomolecular information processing in unicellular organisms

A large number of experimental studies have shown that unicellular organisms possess complex behavior only possible if some kind of biomolecular information is stored and processed.

One of the most studied examples of complex systemic behavior in a single cell is *Physarum polycephalum*. This unicellular species, belonging to the *Amoebozoa* phylum, is an important branch of eukaryote evolution, characterized by a large cytoplasm with many nuclei that remain suspended in a single contiguous protoplasmic volume called plasmodium. The body of the plasmodium contains a complex dynamic network of tubular structures by means of which nutrients and biomolecular elements circulate in an effective manner [Shirakawa and Gunji, 2007].

The movement of the plasmodium is termed shuttle streaming, characterized by the rhythmic back-and-forth flow of biomolecular substances. The dynamically reconfiguring network of tubular structures can redirect the flow of protoplasm towards the plasmatic membrane causing the movement of a mass of flowing pseudopods, and as a result, the organism can crawl over the ground at a speed of approximately 1-5cm/h [Kessler, 1982]. The complex protoplasm exhibits rich spatio-temporal oscillatory behavior, and rhythmically synchronized streams with intracellular oscillatory patterns such as the oscillations in ATP concentration, the plasmagel/plasmasol exchange rhythm and the oscillations in the cytoplasmic free calcium level [Ueda et al., 1986].

In the last several years, a number of studies in *Physarum polycephalum* have shown that this cell is able to store information, learn and recall past events. For example:

- This unicellular organism has the ability to find the minimum-length solution between two points in a maze, which is considered cellular information processing [Nakagaki et al., 2000; Nakagaki et al., 2001]; and it has also been verified that the shortest path problem in the maze is a mathematically rigorous solution [Miyaji and Ohnishi, 2008].
- The *Physarum* constructs appropriate networks for maximizing nutrient uptakes, achieving better network configurations than those based on the shortest connections of Steiner's minimum tree [Nakagaki et al., 2004a; Nakagaki et al., 2004b].

- In particular, high quality solutions for the 8-city travelling salesman problem (TSP), which is known to be NP-hard and hence computationally difficult, have been found by using *Physarum Polycephalum* [Aono et al., 2011a; Aono et al., 2011b; Zhu et al., 2011].
- Other direct experimental studies evidenced that this kind of amoeba can memorize sequences of periodic environmental changes and recall past events during the adaptation to different stimuli (it can even anticipate a previously applied 1 hour cold-dry pattern) [Saigusa et al., 2008].
- Despite being a single multinucleate cell, the plasmodium can be used to control autonomous robots [Tsuda et al., 2007; Gough et al., 2009].
- This cell is able to form an optimized network closely approaching the purposely designed Tokyo railway system [Tero et al., 2010].
- Furthermore, it has also been verified that the plasmodium can solve complex multiobjective foraging problems [Dussutour et al., 2010; Bonner, 2010; Latty and Beekman, 2011].
- Self-organization, self-optimization and resolution of complex optimization problems by means of adaptive network development also occur in the *Physarum Polycephalum* [Marwan, 2010].

All these experimental results push forward the need for an intrinsic memory storage mechanism [Marwan, 2010] and a "*primitive intelligence*" [Nakagaki et al., 2000; Saigusa et al., 2008; Pershin et al., 2009; Nakagaki, 2001; Nakagaki and Guy, 2008].

Many other experimental examples show that other different unicellular organisms have sophisticated behavior including information storage. A pioneer scientist in revealing such behavior was H. S. Jennings, who showed many years ago that a single cell such as Paramecium could have a primitive kind of learning and memory [Jennings, 1905]. Neutrophils also exhibit a rudimentary memory system by which they are able to "recall" past directions [Albrecht and Petty, 1998]. *Dictyostelium* and *Polysphondylium amoebae* seem to have a rudimentary memory and they can show long directional persistence (~10 min), being able to remember the last direction in which they turned [Li et al., 2008]. Even individual neurons seem to display short-term memory, and permanent information can be stored when nerve cells in the brain reorganize and strengthen the connections with one another [Sidiropoulou, 2009].

At a molecular level, information processing [Ausländer et al., 2012; Daniel et al., 2013] has been observed in numerous examples, such as the reversible phosphorylation of proteins [Thomson and Gunawardena, 2009]; the microtubule dynamics [Faber et al., 2006]; the enzymatic processes [Baron et al., 2006; Katz and Privman, 2010]; the redox regulation [Dwivedi and Kemp, 2012]; the transcription processes [Mooney et al., 1998]; the genetic regulatory networks [Qi et al., 2013]; the input signal transductions [Roper, 2007]; the biochemical networks [Bowsher, 2011]; the NF-kappaB dynamics [Tay et al., 2010]; the intracellular signaling reactions [Purvis and Lahav, 2013; Kamimura and Kobayashi, 2012]; the metabolic switches [Ramakrishnan and Bhalla, 2008]; the chemotaxis pathway [Shimizu et al., 2010]; the network motifs [Alon, 2007], and other cellular processes [Ben-Jacob, 2009].

**3.2 Quantitative studies for molecular information processing in self-organized enzymatic sets and in dissipative metabolic networks**

Metabolic patterns may have information which can be captured by the Transfer Entropy (TE) which allows quantifying biomolecular information flows in bits [Schreiber, 2000].

In 2012, the molecular informative properties of a single self-organized multienzymatic set (a metabolic subsystem) were analyzed *in silico* by using Transfer Entropy for the first time [De la Fuente and Cortes, 2012].

According to this analysis, the information patterns contained in the different substrate fluxes are continually integrated and transformed by the set of enzymes that form part of the metabolic subsystem. The TE analysis showed that in the integrative catalytic processes, a dynamic functional structure for molecular information processing emerges, characterized by changing the informative flows that reflect the modulation of the enzymatic kinetic activities. The level of functional influence in terms of causal interaction between the enzymes is not always the same, but varies depending on the information fluxes and the particular dynamic regime of the catalytic system.

The multienzymatic set shapes a functional dynamic structure, able to permanently send biomolecular information between its different enzymatic parts, in such a way that the activity of each enzyme could be considered an information event.

As a result of the overall catalytic process, the enzymatic activities are functionally coordinated, and the metabolic subsystem operates as an information processing system, which at each moment, redefines sets of biochemical instructions that make each enzyme evolve in a particular and precise catalytic pattern.

In terms of information theory, the metabolic subsystem performs three simultaneous functions: signal reception, signal integration, and generation of new molecular information.

In light of this research, the self-organized enzymatic sets seem to represent the most basic units for information processing in the cells [De la Fuente and Cortes, 2012].

On the other hand, when considering the activity of an interrelated set of different metabolic subsystems, a highly parallel, super-complex structure of molecular information processing emerges in the network. This was evidenced in a numerical study of dissipative metabolic networks in which the biomolecular information flows were quantified in bits between the metabolic subsystems by using Transfer Entropy [De la Fuente et al., 2011].

Specifically, the results showed that the network as a whole integrates and transforms the molecular information coming from both the external environment and the individually assembled metabolic subsystems. The basic units of information processing (each self-organized multienzymatic set) work in parallel and coupled with each other, and as a consequence of these collective processes, functional modules, metabolic switches, highly dynamical metabolic synapses and an information processing super-system emerge in the network.

In addition to the topological network structure, characterized by the specific location of each multienzymatic set, molecular substrate fluxes and regulatory signals (mainly allosteric and covalent modulations), a functional network structure of effective connectivity emerges which is dynamic and characterized by having significant variations of biomolecular information flows between the metabolic subsystems.

The dissipative network behaves as a very complex decentralized information processing super-system which generates molecular information flows between the self-organized enzymatic sets, and forces them to be functionally interlocked, i.e., each catalytic set is conditioned to cooperate with others at a precise and specific activity regime in concordance to the activity system as a whole. As a result of the overall functional process, the network defines sets of biochemical instructions at each moment that make every metabolic subsystem evolve with a particular and precise dynamic change of catalytic patterns.

The quantitative analysis also showed that the functional network is able to self-regulate against external perturbations by means of this information processing super-system [De la Fuente et al., 2011].

**3.3.1 Dynamic metabolic memories**

Memory is a fundamental element for the efficient information processing.

Recently, catalytic network dynamics has been numerically analyzed using statistical mechanics tools; in concrete, an equivalent Hopfield network has been obtained using a Boltzmann machine, showing that the enzymatic activities are governed by systemic attractors with the capacity to store functional metabolic patterns. This stored molecular information can be correctly recovered from specific input stimuli [De la Fuente et al., 2013].

In short, the main conclusions of the statistical mechanic analysis were the following:

I. The catalytic dynamics of the dissipative metabolic network are dependent on the Lyapunov function i.e., the energy function of the biochemical system, which is a fundamental element in the regulation of the enzymatic activities.

Starting in any initial state after an external stimulus, and due to the collective enzymatic dynamics, the metabolic network evolves trying to reduce the energy function, which in the absence of noise will decrease in a monotone way until achieving a final system state that is a local minimum of the Lyapunov function. When the biochemical system state reaches a local minimum, it becomes a (locally) stable state for the enzymatic network.

The dissipative metabolic network has multiple stable states, each of which corresponds to a specific global enzymatic pattern of the biochemical system.

II. The enzymatic activities of the metabolic network are governed by systemic attractors, which are locally stable states where all the enzymatic dynamics are stabilized.

The minima of the Lyapunov function corresponds to these emergent systemic attractors, which regulate the metabolic patterns of the subsystems and determine the enzymatic network to operate as an individual, completely integrated system.

The energy function, on which the dynamics of the metabolic network depend, has a complex landscape with multiple attractors (local minima).

III. The emerging systemic attractors correspond with Hopfield-like attractors in which metabolic information patterns can be stored [Hopfield, 1982].

The biochemical information contained in the attractors behaves as a functional metabolic memory i.e., enzymatic activity patterns stored as stable states, which regulate both the permanent catalytic changes in the internal medium, as well as the metabolic responses originated to properly integrate the perturbations coming from the external environment.

When an external stimulus pattern is presented to the system, the network states are driven by the intrinsic enzymatic dynamics towards a determined systemic attractor which corresponds to a set of memorized biochemical patterns.

The formation of the attractor landscape is achieved by metabolic synaptic modifications, i.e., the functional and molecular connections between enzymatic sets, essentially through substrate fluxes, covalent modulations (PTM) and allosteric regulations. These connectivity patterns can be oscillatory or steady states and comprises an infinite number of distinct activity behaviors.

Each biochemical synapse is involved in the storage of the metabolic memory, and the metabolic synaptic plasticity seems to be the basis of the memories stored in the metabolic network.

IV. A key attribute of the analyzed systemic attractors is the presence of "associative memory". When the metabolic network is stimulated by an input pattern, the enzymatic activities converge to the stored pattern, which most closely resembles the input, i.e. the attractors have associative memory.

The metabolic network attractors seem to store functional enzymatic patterns which can be correctly recovered from specific input patterns. The results explicitly show that there are indeed memory attractors and capacity to retrieve information patterns (from the weights) which are local minima of the catalytic activity in the network.

Thus, if the initial condition of the network dynamics starts within the basin of attraction of such a pattern, the dynamics will converge to the originally stored pattern. As it is well known, this class of "pattern-completion" dynamics has been addressed in Hopfield network studies for years, and it is generally named "associative memory" [Amit, 1989].

### 3.3.2 The mean field approximation

In the metabolic memory analysis [De la Fuente et al., 2013], a mean field assumption to describe the equivalent-Hopfield network depicting the metabolic dynamics was computed. The results indicate that by the local minima of the metabolic network energy the information can be stored and eventually retrieved; this is what researchers in neural networks studies described as "having associative memory" [Amit, 1989]. Notice that the original dissipative metabolic network can have much more complicated energy landscapes than the ones approached using the mean field solution.

For more details about information encoding and retrieving (associative memory) and the capacity for pattern storage of a metabolic network see Supplementary Material 01.

**3.4 Post-translational modifications and metabolic memories**

The dynamic of the metabolic synaptic plasticity governed by Hopfield-type attractors is the fundamental element indispensable to store functional metabolic memories in the enzymatic networks.

However, from a bioenergetic point of view, the Hopfield-like metabolic memory is expensive since it requires a permanent activity to maintain the network functionality, expending significant amounts of energy mainly in the form of ATP.

Recent investigations have shown that certain enzymes responsible for post-translational modulation act on multimeric proteins which undergo different modification states that could have slow-relaxation dynamics under specific conditions. The activity of such PTM enzymes depends, in a crucial manner, on the temporal history of the molecular information input they receive, and this information can be stored as a slow relaxation process in which slowness enables long-term maintenance of an embedded state [Hatakeyama and Kaneko, 2014].

This property makes possible an efficient information transfer from the Hopfield-like metabolic dynamics to the enzymes involved in covalent post-translational modulation so that the functional memory can be embedded in multiple stable molecular marks.

The molecular information transfer from the functional metabolic memories to the covalent marks requires a lot less energy than the maintenance of the emergent attractor dynamics in the metabolic networks. Thus, the post-translational modification of proteins seems to represent an essential mechanism for the maintenance of Hopfield-like memories in a structural and energetically efficient way.

PTM is a highly dynamic, essential enzymatic process, which modifies the proteic functionality, reversibly transforming its structure by adding biochemical marks. Consequently, a single protein subjected to PTM may show a broad range of molecular functions and provoke profound physiological effects in cells [Cohen, 2000; Impens et al., 2014].

In fact, PTM is a metabolic mechanism present in all cell types, which can affect the protein structures involved in most cellular processes, e.g., the regulation of the microtubule cytoskeleton [Janke and Bulinski, 2011]; the signal integration [Deribe et al., 2010]; the chromatin regulation [Bannister and Kouzarides, 2011]; the modifications of RNA silencing factors [Heo and Kim, 2009]; the modulation of mitochondrial membrane activities [Distler et al., 2009]; the regulation of different metabolic processes in the nucleus [Gloire and Piette, 2009]; the changes of cytochrome activity [Aguiar et al., 2005]; the regulation of the mitochondrial translation machinery [Koc and Koc, 2012]; and other modulations of metabolic processes [Running, 2014].

There are different types of PTMs, e.g., phosphorylation, methylation, glycosylation, acetylation, ubiquitination, SUMOylation, etc., and the number and abundance of these modifications is constantly being updated [Khoury et al., 2011].

Protein phosphorylation is the most extensively studied PTM, and the phosphorylation-dephosphorylation mechanism is assumed to take place on a fast time scale in comparison to the protein half-lives. Different sites on the same protein can be consecutively phosphorylated-dephosphorylated: these events are not distributed along the whole protein structure but are instead constrained to sites of high accessibility and structural flexibility [Gnad et al., 2007], which allows molecular information encoding, adding or removing molecular marks [Thomson and Gunawardena, 2009].

While bare DNA encodes 2 bits of information per nucleotide, even in a single protein, reversible chemical modifications occur at multiple sites, allowing the encoding of a potentially large amount of information. A protein with $n$ phosphorylated sites has an exponential number ($2^{\wedge}n$) of phospho-forms; the number of phosphorylated sites on a protein shows a significant increase from prokaryotes (with $n \leq 7$ sites) to eukaryotes, with examples having $n \geq 150$ sites, implying that a protein can encode a large amount of molecular information as a function of a varying number of covalent phospho-marks [Thomson and Gunawardena, 2009].

The reversible structural changes in regulatory networks of proteins through PTM allow molecular information encoding and complex information processing, which can originate flexibility and adaptive metabolic responses in the cell [Sims and Reinberg, 2008; Sunyer et al., 2008; Thomson and Gunawardena, 2009].

Attractor network dynamics followed by structural modifications of proteins would represent the two stages of the dynamic memory of cellular metabolism. In accordance with both processes, the metabolic memories are derived from continuous changes in the functionality of the network dynamic (due to metabolic synaptic plasticity) linked to structural molecular modifications (covalent marks).

## 3.5 Examples of post-translational modifications and molecular information storage in the cell

*Escherichia coli* chemotaxis network represents one of the best studied examples of a biochemical dynamic system with the capacity to store functional information using molecular information processing linked to post-translational modification dynamics [Greenfield et al., 2009; Li and Stock 2009; Tu et al., 2008; Shimizu et al., 2010; Stock and Zhang, 2013].

Bacterial cell swimmers integrate environmental information accurately in order to make proper decisions for their own survival, in such a way that they can move towards favorable sites and away from unfavorable environments by changing their swimming patterns. As a result of the chemotaxis system, the signal transduction process translates information into appropriate motor responses and the bacterial cells traverse gradients of chemical attractants displaying directional sensing and moving in an efficient manner by performing temporal comparisons of ligand concentrations. Thus, the cell detects extracellular chemical gradients and senses very small fractional changes, even at nanomolar concentrations [Sourjik and Berg, 2002; Sourjik and Berg, 2004].

Since the pioneering works by Julius Adler in the 1960's [Adler, 1966; Adler and Dahl, 1967; Adler and Templeton, 1967] the biochemical mechanisms that underlie sensory-motor regulation in *E. coli* have been extensively investigated and their principal characteristics have been detailed [Baker et al., 2006; Hazelbauer and Falke, 2008]. Membrane receptors responsible for signal transduction assemble into large clusters of interacting proteins, shaping a complex modular network [Greenfield et al., 2009; Hamadeh et al., 2011; Shimizu et al., 2010] with one main feedback loop [Emonet and Cluzel, 2008; Wadhams and Armitage, 2004]. The network consists of several thousands of alpha-helical transmembrane protein fibers that interact with one another forming a cortical structure below the cytoplasmic membrane. The fiber ends that pass through the cytoplasmic membrane interact with a complex layer of sensory receptor domains.

*E. coli* sensory–motor is functionality regulated by two reversible protein modification processes: phosphorylation and carboxyl methylation. The phosphorylation processes provide a direct connection between sensory receptor complexes and motor responses. The transmembrane protein fibers shape a four-helix bundle with at least eight potentially anionic glutamate side chains that can be either exposed as a minus charge or neutralized by methylation. Each fiber can potentially be in any one of $2^8$ different states of modification, and the increments or decrements in attractant concentration produce behavioral responses that lead to changes in the methylation dynamics [Stock and Zhang, 2013].

The dynamic changes in the chemotaxis network functionality provoke specific molecular information processing and, as a consequence, structural molecular modifications in form of methylations, and they are carried out in such a way that the cell can record recent chemical past by using these reversible methylation in determined glutamic acid residues [Li and Stock, 2009].

The dynamic methylation-demethylation patterns linked to information processing serve as a structural functional memory allowing the detection of attractant gradients by comparing current concentrations to those encountered in the past. The carboxyl methylation mechanism stores information concerning the environmental conditions that the bacterium has experienced, and these dynamic patterns of glutamyl modifications act as a structural dynamic memory which allows cells to respond efficiently to continual changes in attractant concentrations [Greenfield et al., 2009; Li and Stock, 2009; Stock and Zhang, 2013].

*E. coli* chemotaxis network exhibits a rich variety of behavior and dynamic properties such as robustness [Kollmann et al., 2005; Alon et al., 1999; von Dassow et al., 2000], signal amplification [Tu, 2013], molecular information processing [Shimizu et al., 2010], fast response and ultra-sensitive adaptation [Bray et al., 1998; Kollmann et al., 2005; Wadhams and Armitage, 2004; Sourjik, 2004].

The essential components of the *E. coli* chemotaxis system are highly conserved among all motile prokaryotes [Wuichet et al., 2007]. However, many species have chemotaxis networks that are much more complex than that of *E. coli* [Hamadeh et al., 2011],

Although the chemistry of signal transduction in bacterial and vertebrate cells differs substantially, there are numerous fundamental similarities [Stock et al., 2002]. As in prokaryotes, signal transduction in eukaryotic cells generally involves phosphotransfer mechanisms and other post-translational modifications that couple metabolic energy to information processing [Li and Stock, 2009].

In particular, the biochemical systems of molecular information processing linked to post-translational modification dynamics are highly evolved in neuronal cells.

Neurons are connected through synapses that undergo long-lasting activity dependent alterations in their transmission efficacy, permitting learning and response to experience. One of the principal means to control synaptic transmission is regulation of neurotransmitter receptors, especially the AMPA-type glutamate receptors (AMPARs) that represent the main excitatory neurotransmitter receptors in the neurons. The number of AMPA receptors at the synapse is extremely dynamic and strictly regulated through the addition or removal of receptors resulting in long-term synaptic potentiation or depression (LTP or LTD, respectively) [Anggono and Huganir, 2012]. These synaptic plasticity events are thought to underlie information storage. In particular, reversible post-translational modifications of AMPARs, including phosphorylation [Wang et al. 2005], palmitoylation [Hayashi et al., 2005], ubiquitination [Schwarz et al., 2010], SUMOylation [Luo et al., 2013], N-glycosylation [Tucholski et al., 2014] and O-GlcNAcylation [Taylor et al., 2014] are important regulatory mechanisms of synaptic AMPA receptor expression and function.

By controlling the expression and function of AMPARs, these multiple forms of post-translational modifications of AMPA receptors vigorously regulate plasticity events and many important functions linked to molecular information processing, including learning and memory [Takeuchi et al., 2014].

While prevailing models of long-lasting information storage identify mRNA translation as essential for learning and memory in neural cells, there is evidence that memory storage at PTM level can occur in the absence of protein synthesis [Routtenberg and Rekart, 2005; Routtenberg, 2008].

## 3.6 Long-term and short-term memory

Since the beginning of the neuronal network modeling of associative memory, the connectivity matrix in Hopfield network was assumed to result from a long-term memory learning process, occurring in a much slower time scale than the neuronal dynamics [Willshaw et al., 1969; Willshaw et al., 1970; Amari, 1975; Hopfield, 1982; Amit 1989; Hertz et al., 1991]. Therefore, it is well accepted that the attractors emerging in the neuronal dynamics described by Hopfield networks are the result of a long-term memory process.

Besides, long physiological recordings of neuronal processes exhibit some type of memory structure. Among others, there are at least two commonly used indicators for this kind of memory: the Hurst exponent [Das et al., 2003; De la Fuente et al., 2006] and the presence of long range correlations in the plasticity dynamics for synaptic weights- measured, for instance, with long tails in the synaptic distribution of weights [Barbour et al., 2007]. Several phenomenological models have shown that this persistence in the dynamics of weights underlying short-term memory, eventually might be translated into a stable long-term memory after memory consolidation [Fusi et al 2005; Mongillo et al., 2008; Itskov et al., 2011; Romani et al., 2013], but the precise mechanisms making this transition possible are not yet well understood.

Long-term correlations (mimicking short-term memory in neuronal systems) have also been analyzed in different metabolic processes not belonging to neuronal cells. One of the most studied is the calcium-activated potassium channels which has been observed in Leydig cells [Varanda et al., 2000; Bandeira et al., 2008], kidney Vero cells [Kazachenko et al., 2007], and human bronchial epithelial cells [Wawrzkiewicz et al., 2012]; moreover, other biochemical processes also present long-term correlation, for instance, the intracellular transport pathway of *Chlamydomonas reinhardtii* [Ludington et al., 2013], the NADPH series of mouse liver cells [Ramanujan et al., 2006], the mitochondrial membrane potential of cardiomyocytes [Aon et al., 2008], and different biochemical numerical studies on enzymatic pathways and dissipative metabolic networks [De la Fuente et al., 1998a; De la Fuente et al., 1998b; De la Fuente et al., 1999].

These studies support that, at the molecular level, metabolic memories seem to exhibit both long-term and short-term memory, but this issue requires further research.

# 4. The System CMS-DNA

## 4.1 Elements of the Cellular Metabolic Structure

Metabolism is the largest known source of complexity in nature.

The cellular enzymatic processes shape a highly structured dynamic super-system with different levels of organization, which have a large number of adaptive molecular mechanisms acquired throughout biological evolution.

On the way to structural and functional complexity, metabolic evolution has given rise to a great diversity of biochemical organizational forms, ranging from the sophisticated bacterial metabolisms, through the complex enzymatic networks belonging to protists, to the extraordinary structures and processes derived from multicellular eukaryotic organizations such as embryogenesis or the neural networks of higher mammals.

The millions of living metabolic species represent the inexhaustible source of complexity developed by the dynamic forces that structure the metabolic systems.

Enzymes are the fundamental molecules for metabolic life. At a basic level, they can shape functional associations dissipativelly structured (metabolic subsystems), which seem to constitute elemental catalytic units of the cell reactor (Section One).

A rich variety of dynamic catalytic patterns emerge in each metabolic subsystem (including oscillatory and quasi-steady states), which correspond to distinct activity regimes (Section One). Catalytic changes allow the metabolic synapses, i.e., the structural and functional dynamic connections between enzymatic sets. As a result of the flexible activity regimes, flux plasticity and structural plasticity also appear in the cell (Section Two).

At a superior level, enzymatic processes are organized into distinct modular metabolic networks, which work with relative independence of the global system. In turn, the whole metabolism seems to be functionally integrated, generating systemic responses from the concerted actions of modular networks, shaping a complex dynamic super-system: the Cellular Metabolic Structure (CMS in short). Therefore, the two principles that link the main modes of metabolic connectivity in the cell are functional segregation and functional systemic integration (Section Two).

The CMS is an integrated catalytic entity that acts as an information processing system capable of storing functional biochemical patterns. The metabolic synaptic dynamics governed by Hopfield-type attractors and the structural covalent modifications represent the two stages of the dynamic memory of cellular metabolism. Plasticity seems to be the key essential property for the emergence of metabolic memory (Section Three).

All these properties constitute some of the main elemental principles of the CMS, which allow the self-organization, self-regulation and adaptation of the cell to the external medium.

**4.2 Self-regulation of the Cellular Metabolic Structure**

Unicellular organisms are open metabolic systems.

To maintain the self-organized biochemical functionally, the CMS needs a permanent exchange of energy and matter with the external medium. As a consequence, metabolic life would not be possible without it having the capacity to adequately respond to the large possible combinations of external variables to which it is exposed.

The CMS is a sophisticated, sensitive system, whose enzymatic processes change immediately in response to any variations in the environmental conditions [Almaas et al., 2004; Almaas et al., 2005; Papp et al., 2004; Wang and Zhang 2009].

When the external perturbations are small, the CMS changes accordingly, and slowly reverts back to the pre-stimulus dynamic by reorganizing and adjusting the catalytic patterns and metabolic fluxes [Inoue and Kaneko 2013].

If the external perturbations are stronger, i.e., the replacement of a major source of nutrients for another, the CMS computes the various possible ways to efficiently process the molecular substrates, turning off certain metabolic activities (even completely eliminating specific enzymatic sets), and adding new reactive processes and alternative routes. As a result, the metabolism shows extensive reorganization in the metabolic functionality resulting in a rewiring of the global network fluxes [Storlien et al., 2004; Buescher et al., 2012; San Román et al., 2014; Zhang et al., 2014].

When the metabolic system undergoes extreme external conditions as, for instance, nutrient deprivation, the CME generates very complex enzymatic reorganization comprising the entire system. In general, the autophagy interaction network is activated in response to starvation [Behrends et al., 2010], and as a consequence, proteins, organelles and other cellular substructures are sequestered in autophagosomal vesicles to be later degraded. This is one of the major mechanisms by which a starving cell reallocates nutrients from unnecessary metabolic pathways to more-essential processes [Shintani and Klionsky, 2004].

Under starvation conditions, the CMS computes the efficient rates of molecular turnover, modulates the systemic enzymatic processes with extreme precision, and adjusts the number of different organelles to match the available resources, e.g., the metabolic system drastically regulates different mechanisms such as the ribosome biosynthesis and degradation [Pestov and Shcherbik, 2012], the nucleolar proteome dynamics [Andersen et al., 2005], the protein turnover [Neher et al., 2003]; the lipid degradation [Holdsworth and Ratledge 1988], the peroxisomal number [Mijaljica et al., 2006], and the global cellular energy [Boender et al., 2011] among others.

There are endless possible combinations of environmental variables (physical, chemical, nutritional and biological) to which the cell must give an answer. The CMS has the properties to self-adapt against these numberless external variables, displaying an impressive capacity to regulate thousands of metabolic processes simultaneously, computing the appropriate regulatory responses at each time through complex and specific catalytic dynamic patterns that must properly correspond to each set of external perturbations.

Furthermore, these self-regulations occur very quickly, for instance, the metabolic dynamics of the Escherichia coli were studied under cultures in which the glucose concentration was constantly changing, following an up-and-down pattern. These studies show that the systemic energetic state of the cell (measured by the adenylate energy charge) self-adjusts as a whole in only 2 min after the external perturbations [Weber et al., 2005].

In addition, the metabolism also exhibits complex and continuous cycles of molecular turnover which includes the complete proteome, glycome, lypidome, metabolome and transcriptome of the cell. More specifically, enzymes, the essential molecular actors for cell functionality, are continually synthesized and destroyed in a dynamic turnover process in which all proteins are broken down partially, into peptides, or completely, into amino acids for later being regenerated de novo [Doherty et al., 2009; Li, 2010]. Even when a cell is not undergoing growth, the protein pool is dynamic and the proteolytic machinery continues its relentless process of molecular turnover, requiring for it significant energy [Doherty and Beynon, 2006; Doherty et al., 2009].

As a result of the whole metabolic dynamic that encompasses reactive molecular transformations including protein turnover, determined polypeptides must be suitably synthetized when the physiological conditions require it. Therefore an adequate molecular information flow from the CMS to the DNA is essential for ensuring the adequate polypeptide replacement.

## 4.3 Nexus between the CMS and the DNA

Just over 40 years ago, the Nobel Prize John Gurdon showed the existence of a molecular flow from the cytoplasm into the nucleus involved in the differential expression of certain genes during early embryonic development [Gurdon, 1971]. The experiments were realized transplanting the nucleus of a cell, which was carrying out one specific dynamic activity, into an enucleated cytoplasm whose metabolic activity was quite another. These studies were an extension of Briggs and King's works on transplanting nuclei from embryonic blastula cells [Briggs and King, 1959].

In the aforementioned Gurdon's experiments, a fundamental conclusion was reached: some cytoplasmic molecules control the expression of certain genes (or their repression) in an independent way. In all hybrid cells, the transplanted nucleus did not continue with its previous activity, but changed their function conforming to that of the host cytoplasmic metabolism to which it had been exposed.

For instance: the mid-blastula nucleus synthesizes DNA but not RNA, and when it is implanted into the cytoplasm of a growing oocyte whose nucleus does not synthesize DNA, but RNA, the new hybrid cell stops synthesizing DNA and starts synthesizing RNA; if this same mid-blastula nucleus is injected into the cytoplasm of a mature oocyte which did not synthesize any nucleic acid and whose chromosomes are placed in spindle phase, the implanted nucleus ends up synthesizing DNA and the chromosomes arrange in the spindle of cell division; and when an adult neuron nucleus, which synthesizes RNA but rarely synthesizes DNA, is injected into an unfertilized but activated cytoplasm egg whose nucleus would normally synthesize DNA but not RNA, the transplanted nucleus stops the RNA synthesis and begins DNA synthesis [Gurdon, 1971].

Another one of Gurdon's experimental series proved that certain cytoplasmic components select which genes will be activated and/or repressed at any one time; and consequently, it was observed that major changes in different kinds of gene activity respond by molecular signals came from the cytoplasmic metabolism [Gurdon, 1971].

Analogous experiments have been conducted in protozoans, amphibians, plants and in different mammalians such as cats, mice, monkeys rabbits, horses, donkeys, pigs, goats and cows; even a mouse was cloned using the nucleus of an olfactory neuron [Eggan et al., 2004]. Of all these, the most popular example is Dolly, that was the result of an experiment in which a single nucleus come from a differenced mammary cell was successfully fused to an enucleated unfertilized egg from another adult sheep [Wilmut et al., 1997].

According to the current vision, the DNA should have all the essential information for the maintenance and development of the cell, and therefore the mammary nucleus should modify the metabolic activity of the cytoplasm in which it has been injected, so that the new hybrid organism should act as a mammary cell. But what happens is just the opposite. It is the transferred nucleus that undergoes a radical transformation, so that the molecular information flux originated from the host cytoplasmic metabolism reprograms the genetic expression of the implanted nucleus expressing transcription patterns completely different from the previous non-transplanted conditions, and this new program of molecular instructions comes from the metabolic structure of the host cytoplasm.

The gene expression patterns are regulated by the cytoplasmic metabolic structure in which the mammary nucleus was injected, and no longer behaves like in a mammary cell but starts dividing according to the new molecular program (sequence of instructions to perform a specified biochemical task) executed by the cytoplasmic metabolism. During the first three divisions the new cell replicates its DNA without expressing any of the genes, and it is followed by successive cell divisions and progressive differentiation, first into the early embryo and subsequently into all of the cell types that characterize the animal.

Cellular Metabolic Structure (the cytoplasm and nucleoplasm metabolism as a whole) is able to reprogram the gene expression of mammary DNA, and both the CMS and the DNA coordinate together the development of an entire organism.

In eukaryotic cells, there is a long history of experimental observations about the molecular flow from the cytoplasmic metabolism to the nucleus, dating back to 60s of the last century [Zetterberg, 1966; Arms, 1968; Merriam, 1969; Gurdon and Weir, 1969] to the present [Cook et al., 2007; Wimmer et al., 2013; Tamura and Hara-Nishimura, 2014].

A great variety of experimental studies have shown that the molecular flow from the cytoplasmic metabolism to the nucleoplasm consists of ordinary components of the cellular metabolism. Thousands of proteins, metabolites, cofactors and ions are transferred to the nucleoplasm with remarkable efficiency and specificity [Kuersten et al., 2001; Cook et al., 2007; Gerhäuser, 2012], and the different molecular combinations (with precise concentrations) allows for modulation of the DNA-associated metabolism (essentially integrated in the transcriptional system), which regulates the gene activity.

Extensive experimental studies both in prokaryotes and eukaryotes have shown the same fundamental process: genes exhibiting transcriptional induction in response to metabolic changes, originated during self-adaptation to a wide range of external factors such as nutrient deprivation, chemical agents (the hormones, the pH, the ionic strength, and others) and physical fluctuations (including temperature shock, drought, high-intensity light etc.) [Causton et al., 2001; Gasch and Werner-Washburne, 2002; Kreps et al., 2002; Hazen et al., 2003; Liu et al., 2005; Deutscher et al., 2006; Stern et al., 2007; Swindell et al., 2007; Wilson and Nierhaus, 2007; Angers et al., 2010; Mirouze and Paszkowski, 2011]. For instance, under glucose starvation, the stringent reduction of ATP turnover at the metabolic level was accompanied by a strong down-regulation of genes involved in protein synthesis [Boender et al., 2011].

The traffic from the CMS to the DNA-associated metabolism ensures the intimate subordination of the transcriptional system to the needs of the metabolic activity, at each time.

The transcriptional system seems to be one the most complex dynamic structures of the cell. It is formed by different, highly coordinated metabolic networks which integrate a great variety of signals and process molecular information from both the DNA and the CMS.

Unfortunately, most of the dynamic regulatory aspects of the metabolic transcriptional system are still poorly understood.

A surprising variety of biochemical mechanisms participate in the control of gene expression, including the topoisomerases, the transcription factors, the RNA polymerases, the small RNAs, the translation system, the turnover rate of proteins, the dynamic levels of the adenylate energy charge, etc. These metabolic processes are in turn modulated by complex molecular networks [Balázsi et al., 2005; Missiuro et al., 2009], already described in the miRNAs [Dong et al., 2010; Gennarino et al., 2012; Schulz et al., 2013], the transcription factors [Walhout, 2006; Neph et al., 2012], the RNA polymerase complexes [Coulombe et al., 2006; Schubert, 2014], and the mRNA dynamics [Maquat and Gong, 2009; Schott and Stoecklin, 2010; Vlasova-St Louis and Bohjanen, 2011] among others.

Chromosome packaging is another essential biochemical control process for gene expression, common to all living organisms. In eukaryotes, the nucleosome-based DNA organization allows reversible access of different metabolic regulatory mechanisms in order to modulate determined DNA sequences. In fact, the tails of the four core histones are subject to a variety of enzyme-catalyzed, post-translational modifications [Turner, 2005]. Networks of biochemical factors that functionally interact with the nucleosome, reversibly remodeling it, have also been observed [Burgio et al., 2008].

Prokaryotic cells, despite exhibiting a very different DNA organization than that of the eukaryotes, also present a protein collectivity referred to as nucleoid-associated proteins, which contribute to the compaction and regulation of the genome [Rimsky and Travers, 2011; Takeda et al., 2011].

In addition, gene expression can also be modulated through the methylation-demethylation dynamics on the DNA nucleotides, mostly at CpG sites to convert cytosine to 5-methylcytosine. The enzymatic systems for these covalent marks are also integrated in the DNA-associated metabolism, and seem to be directly involved in maintaining the dynamic metabolic memories in the transcriptional system.

In brief, both eukaryotes and prokaryotes appear to have a common mechanism to coordinate the two information storage systems in the cell: the CMS and the DNA (Figure 1b). The CMS integrates and process the stimuli coming from the environment, generating self-regulatory metabolic responses and a molecular information traffic to the DNA-associated metabolism. This information flux ensures the intimate integration and subordination of the transcriptional system to the rest of metabolic activity of the cell. The molecular flow allows the appropriate orchestration of the whole transcriptional system and, as a consequence, determined parts of the genome are activated and deactivated at strategic times; so that, in accordance to the requirements of the adaptive maintenance of the cellular metabolism at each moment, gene expression is accurately regulated and only specific polypeptides are synthesized.

Other important implications of the CMS in the regulation of DNA-associated metabolism such as the generation of new polypeptidic information by means of alternative splicing or directed mutations are beyond the scope of the present work.

**4.4 Duality of the cellular hereditary information.**

In addition to the molecular DNA-based inheritance, a wide variety of experimental evidences have shown that determined functional metabolic processes codified via covalent mark patterns, not encoded in the DNA sequence (epigenetic changes), can be preserved when cells divide and thus can be transferred to the next generation.

For instance, epigenetic changes originated by environmental impacts on the cellular metabolism have been observed to extend over the offspring in different kinds of cells [Pembrey et al., 2006; Gluckman et al., 2007; Nadeau, 2009; Jiminez-Chirallon et al., 2009; Carone et al., 2010; Grossniklaus et al., 2011].

Nutrition also may induce specific metabolic processes that seem to be transmitted to the next generation by means of covalent marks that do not involve changes in the underlying DNA sequence of the cell [Kaati et al., 2007; Waterland et al., 2007].

Besides, experimental findings in cells under stress shed new light on transgenerational inheritance of metabolic memory processes [Seong et al., 2011; Crews et al., 2012; Pecinka and Mittelsten Scheid, 2012].

Likewise, persistent and stable transmission of epigenetic molecular information across generations has been described. For instance, it was reported in *Arabidopsis thaliana* that environmental stress leads to a raise in genomic flexibility (increasing the hyper-recombination state), which persisted in a stable way for up to four untreated generations [Molinier et al., 2006].

Another example of stably inherited DNA methylation pattern occurs in a variant of *Linaria vulgaris* originally described more than 260 years ago by Linnaeus [Linnaeus, 1749], in which the fundamental symmetry of the flower is changed from bilateral to radial as a result of extensive methylation of the *Lcyc* gene that controls the formation of dorsal petals [Cubas et al., 1999].

The complex reprogramming of the metabolic information marks in primordial germ cells according to their parental origin is another example of notable relevance of heritable biochemical patterns not caused by changes in the DNA sequence [Seisenberger et al., 2012; Hanna and Kelsey, 2014].

Moreover, the transmission across generations of epigenetic metabolic information has been observed in bacteria [Casadesús and Low, 2013; Gonzalez et al., 2014].

Epigenetic processes seem to be the manifestation of cellular metabolic memory. The covalent modification of histones and the methylation-demethylation dynamics on specific DNA nucleotides appear to represent the storage of determined functional biochemical memories in a structural way and these marks could only emerge due to the Hopfield-type metabolic dynamics.

There is enough evidence supporting that the unicellular organisms can exhibit two storage information systems capable of transmitting determined information to next generations: the CMS and the DNA. The molecular information of both systems seems to coordinate the physiological development of the whole cell.

## 5. Concluding remarks

The observations and consequent reflections I have made concerning numerous cellular processes and different quantitative biomolecular studies, both presumably interrelated, have led me to consider, without underestimating the enormous importance of DNA, that the elements responsible for the organization and regulation of the fundamental biological processes can be found in determined properties and characteristics of the cellular metabolism.

As far as I have been able to ascertain, I find it probable that the essential attributes upon which cellular functionality lies have their basis in the as yet insufficiently understood *Cellular Metabolic Structure*.

Throughout these lines I have tried, in a conceptual manner, to present the basic elements of a dualist framework concerning the information stored in the cell, according to which the dynamic metabolic memory and the genetic memory seem to be the manifestation of two systems of information storage in the cell. From this perspective, rather than a *genoteque* in evolution, the cell could be considered, a singular metabolic entity, able to permanently self-organize and self-regulate as well as adapt to the outside medium.

The work presented is necessarily and predominantly an exercise of multidisciplinary integration, where concepts from different branches of biological knowledge –mainly Biochemistry, Physiology, Cellular Biology and Molecular Biology– are addressed within the approach of Systems Biology. Moreover, during the development and completion of the quantitative investigations, it has been necessary to study many experimental and numerical data through the use of different analytic systems such as differential calculus, statistical mechanics, discrete mathematics, computing techniques and artificial intelligence, as well as other physico-mathematical tools.

However, in this manuscript, any physico-mathematical formulation has been omitted in order to favour the didactic character of the text. That being said, it is necessary to stress that, in my opinion, without mathematics –the language of nature– it would not be possible to understand precisely the metabolic dynamics and the catalytic functionality that sustain them.

Fortunately, today the necessary application of mathematics to all fields of human knowledge, biology in particular, is widely accepted. In the autobiographical recollections that Charles Darwin wrote in 1876 for his children, he expressed the importance of understanding "something of the great leading principles of mathematics, for men thus endowed seem to have an extra sense" [Darwin, 1905].

This novel dualist framework of functional and structural information stored in the cell could open new perspectives on the comprehension of subjects of notable interest as yet not well understood, such as cellular differentiation, to name but one. Thus, cells with a different *Cellular Metabolic Structure* and, therefore, different catalytic activity programs, but with the same genetic heritage, can express different genes and manifest different physiological behaviour. Something along these lines could be said about the Theory of Evolution, in terms of how it was presented by the endearing Charles Darwin, and the possibility that the cells themselves are able to self-adapt to the medium permanently through the dynamic properties of cellular metabolism and the dualist character of cellular hereditary information.

Within the synopsis presented in these pages, no doubt some of opinions may appear rushed and the reflections I have presented, insufficient.

Frankly, I have to admit that dealing with these overwhelmingly complex issues of systemic cell biology with the desired clarity and precision is extraordinarily difficult. In addition, at this time our comprehension of the dynamic biological phenomena in its most elemental aspects is still very incomplete and insufficient.

It is truly extraordinary that in such a reduced space, of merely a few microns, millions of biochemical processes form a metabolic entity that is sensitive, self-organized and self-regulated, able to process and store information, perpetuating itself in time.

A minute place in the cosmos, where millions upon millions of atoms, formed at the very origin of the initial singularity and in the stars, self-organize to give birth to the molecules of highest complexity and information value in all the known universe.

**Fig 1**

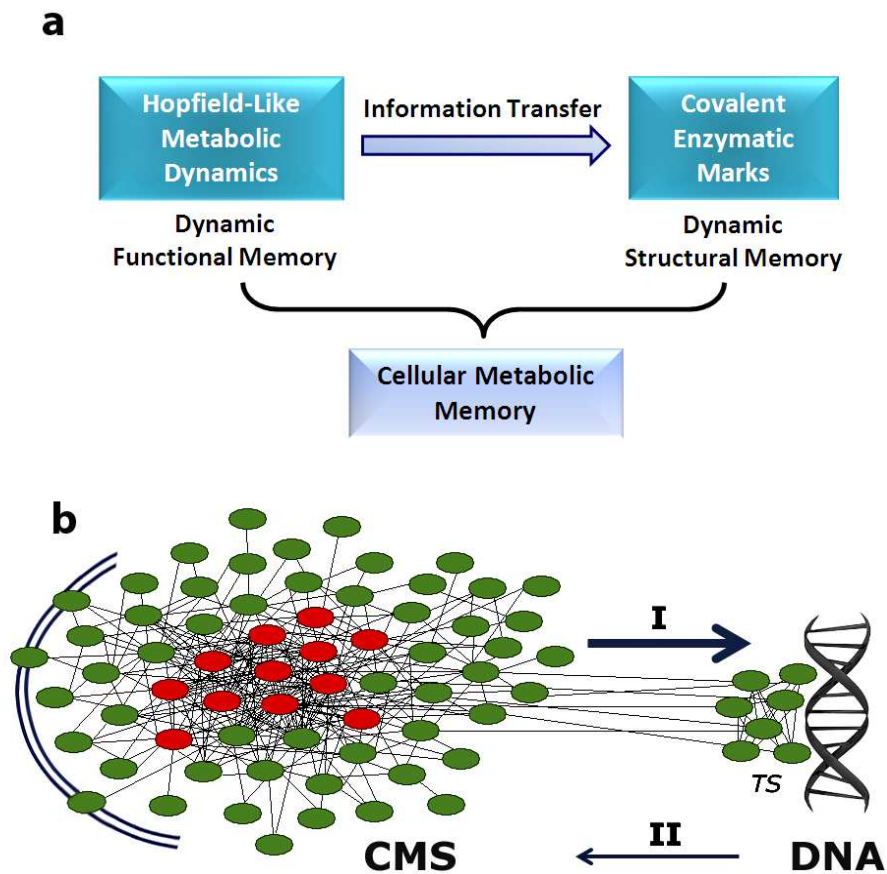

**Figure1. Cellular metabolic memory**
(**a**) Hopfield-like attractor dynamics can emerge in the metabolic networks. They have the capacity to store functional catalytic patterns which can be correctly recovered by specific input stimuli. These metabolic dynamics are stable and can be maintained as long-term memory. In addition, specific molecular information can be transferred from the functional dynamics of the metabolic networks to the enzymatic activity involved in the covalent post-translational modulation so that specific functional memory can be embedded in multiple stable molecular marks. Both the metabolic dynamics governed by Hopfield-type attractors (functional processes) and the enzymatic covalent modifications of determined molecules (structural dynamic processes) seem to represent the two stages of the dynamic memory of cellular metabolism.
(**b**) Between the extracellular environment and the DNA, there appears to exist a cellular metabolic structure (CMS) that behaves as a very complex decentralized information processing system with the capacity to store metabolic memory (Figure 1a). The CMS generates sets of biochemical instructions that make each enzymatic activity evolve with a particular and precise dynamic of change, allowing self-regulation and adaptation to the external medium. In addition, the CMS permanently sends a flow of molecular signals to the DNA-associated metabolism, which shapes the complex transcriptional system. The molecular flows allow accurate regulation of gene expression, so that only specific polypeptides that are required for adaptive maintenance of the CMS are synthesized. Together, both informative systems (CMS and DNA) coordinate the physiological development of the cell. I: determinative molecular flux for the regulation of the transcriptional system. II: polypeptidic reposition. *TS*: transcriptional system. The CMS is represented by a network. Red circles: metabolic core. Green circles: on-off metabolic processes.